\documentclass[journal]{IEEEtran}
\usepackage{graphicx}

\begin{document}
\title{HX-POL -- A Balloon-Borne Hard X-Ray Polarimeter}
%
%
\author{H.\ Krawczynski$^1$, A.\ Garson III$^1$, J.\ Martin$^1$, Q.\ Li$^1$, M.\ Beilicke$^1$, P.\ Dowkontt$^1$, K.\ Lee$^1$, E.\ Wulf$^2$, J.\ Kurfess$^3$, E.\ I.\ Novikova$^2$, G.\  De Geronimo$^4$, 
M.\ G.\ Baring$^5$, A.\ K.\ Harding$^6$, J.\ Grindlay$^7$, J.\ S.\ Hong$^7$\\[2ex]
{\small 
(1) Washington University in St. Louis and McDonnel Center for the Space Sciences,  St. Louis, MO,
(2) High-Energy Space Environment Branch, Naval Research Laboratory, Washington, DC, USA, 
(3) Praxis, Inc., Alexandria, VA, USA,
(4) Instrumentation Division, Brookhaven National Laboratory, Upton, NY, USA,
(5) Department of Physics \& Astronomy, Rice University, Houston, TX, USA,
(6) Astrophysics Science Division, NASA Goddard Space Flight Center, Greenbelt, MD, USA,
(7) Harvard-Smithsonian Center for Astrophysics, Cambridge, MA, USA.}}

\maketitle
\begin{abstract}
We report on the design and estimated performance of a balloon-borne hard X-ray polarimeter called HX-POL. 
The experiment uses a combination of Si and Cadmium Zinc Telluride detectors to measure the polarization 
of 50 keV-400 keV X-rays from cosmic sources through the dependence of the angular distribution of Compton scattered 
photons on the polarization direction. On a one-day balloon flight, HX-POL would allow us to measure the polarization 
of bright Crab-like sources for polarization degrees well below 10\%. On a longer (15-30 day) flight from Australia or 
Antarctica, HX-POL would be be able to measure the polarization of bright galactic X-ray sources down to polarization 
degrees of a few percent. Hard X-ray polarization measurements provide unique venues for the study of particle acceleration 
processes by compact objects and relativistic outflows. In this paper, we discuss the overall instrument design and 
performance. Furthermore, we present results from laboratory tests of the Si and CZT detectors.
\end{abstract}
\section{Introduction}
\IEEEPARstart{X}{-ray} astronomy has made major contributions to modern astronomy and cosomology for the 
last three to four decades. To name a few examples, X-rays allowed us to study the hot gas in which the galaxies 
of galaxy clusters are embedded, to study gas briefly before it plunges into stellar mass and supermassive black holes, 
and to study mass accreting neutron stars. Presently, the National Aeronautics and Space Administration (NASA) and the 
European Space Agency (ESA) fly the Chandra and XMM-Newton soft X-ray telescopes, which set world records in terms of 
angular resolution (Chandra: 0.5'') and collection area (XMM-Newton: 4,300 cm$^2$ at 1.5 keV).
Even though X-ray astronomy has been extremely successful over several decades, there are still largely
unexplored areas which future space-borne X-ray telescopes can pioneer. The NuSTAR hard X-ray telescope\footnote{http:$//$www.nustar.caltech.edu$/$}
is scheduled for launch in 2011 and will image 6-80 keV X-rays with an angular resolution of 40''. 
The International X-ray Observatory (IXO)\footnote{http:$//$ixo.gsfc.nasa.gov$/$} would enable high-throughput excellent energy 
resolution (E/$\Delta$E$>$2400 at 6 keV) X-ray spectroscopy in the 0.3-10 keV energy band. The EXIST 
black hole finder probe\footnote{http:$//$exist.gsfc.nasa.gov$/$} would use coded mask imaging to scrutinize the entire sky every 170 min in 
the 5 keV to 600 keV energy band. 

This paper focuses on X-ray polarimetry, another very promising and largely unexplored area. 
X-ray polarimetry would increase the parameter space for the study of compact objects like black holes and neutron stars 
from two dimensions (time variability and energy spectra) to four dimensions, by adding two qualitatively new parameters: polarization 
degree and polarization direction \cite{Rees:75,Shap:75,Shap:76,Mesz:88}. 
The polarization measurements can be used to identify the processes responsible for the observed X-ray
emission, to constrain viewing perspectives and opacities in accretion disk systems, and to gain access 
to the properties of the magnetic field (order and orientation) in the emission regions.
Recently, concepts have been developed to measure the polarization of soft X-rays by tracking the photo-effect electrons 
that tend to be ejected parallel to the electric field vector of the X-rays with gas pixel 
detectors \cite{2008arXiv0810.2700C} and with time projection chambers \cite{2007SPIE.6686E..29H}. 
At $>$20 keV energies, the Compton effect can be used to measure the polarization, as photons are 
preferentially scattered in the direction perpendicular to the electric field vector.
Early soft X-ray polarization measurements were done with the OSO~8 experiment which used mosaic crystals 
of graphite to yield polarization-sensitive Bragg reflection of 2.6 keV and 5.2 keV X-rays \cite{Weis:78}. 
The Ge detectors on board the INTEGRAL $\gamma$-ray observatory were used to measure the polarization of 
gamma-rays based on Compton effect polarimetry \cite{2008Sci...321.1183D}. Recently, NASA approved the 
Gravity and Extreme Magnetism SMEX (GEMS) mission as a ``small explorer mission''. GEMS will combine 
grazing incident X-ray mirrors with time projection chamber detectors to measure 
the polarization of 2-10 keV X-rays, and will be launched in the next decade \cite{GEMS}. 
The japanese ASTRO-H mission (formerly called ``NeXT'') will have a ``soft $\gamma$-ray telescope'' which uses Si and CdTe 
detector stacks and will have polarization sensitivity in the $\gamma$-ray energy regime \cite{next}.  

\begin{figure}[t]
\begin{center}
\includegraphics[angle=-90,width=3.5in]{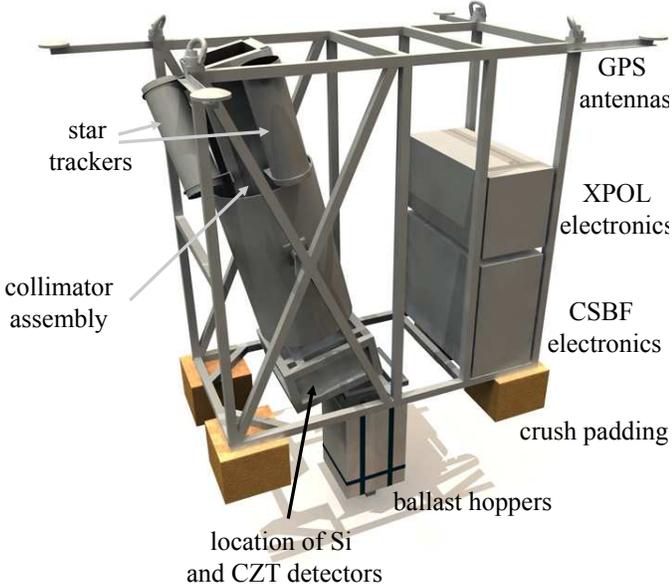}
\vspace*{0.3cm}
\caption{\label{gondola} Conceptual design of the {HX-POL} balloon payload. 
The polarization is measured with a Si-CZT detector configuration inside a 2~m high collimator-shield assembly. 
The collimator is pointed towards a source by rotating the gondola and by changing the elevation.}
\end{center}
\end{figure}

The OSO~8 observations revealed a polarization degree of $\sim$20\%  of the 2.6~keV and 5.2~keV X-ray emission from the 
Crab Nebula, and a polarization angle aligned around 30 degrees oblique to the X-ray jet \cite{Weis:78}. 
In the 0.1-1 MeV energy range, the INTEGRAL observations showed a high polarization 
degree of 46\%$\pm$10\%, and a polarization direction aligned with the orientation 
of the X-ray jet \cite{2008Sci...321.1183D}. The result indicates that the X-ray 
emission comes from the inner jet. The fact that the polarization angle differs from the classic X-ray
band to the soft gamma-ray band provides a significant diagnostic on the jet-nebula environment.
It may be a general trend that polarization fractions increase with the energy of the X-rays, 
as the electrons emitting higher-energy X-rays lose their energy faster than the electrons
responsible for the lower-energy X-rays; as a consequence, the regions responsible for harder X-rays 
should be more compact and thus more uniform than the regions responsible for soft X-rays.
A more uniform magnetic field would lead to a higher degree of polarization.
However, such trends would critically depend on the relative order of the magnetic field on different 
spatial scales, a salient property for MHD and plasma models.
Even at balloon altitudes of $\sim$40 km, the atmosphere is only transparent for $>$20 keV X-rays.
Soft X-ray polarimeters have thus to be space borne. In the hard X-ray energy band, balloon-borne 
experiments with $\sim$500 cm$^2$ detection areas can measure polarization fractions of between 
10\% and a few percent for bright, mostly galactic sources with fluxes between 10\% and 100\% of the flux from the Crab Nebula.
The measurement of the hard X-ray polarization of very short transient events (Gamma Ray Bursts, GRBs) and
most extragalactic sources requires larger, most likely, space-borne experiments.
\begin{figure}[t]
\begin{center}
\vspace*{2cm}
\includegraphics[angle=+90,width=3.5in]{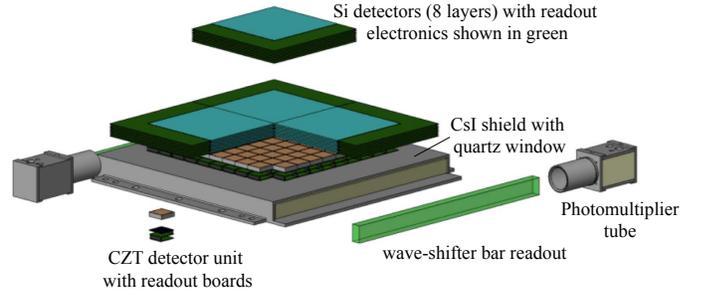}
\vspace*{2.26cm}
\caption{\label{SiCZT} ``Exploded view'' of the {HX-POL} detector assembly. A total of 32 Si detectors 
(each 0.2~cm$\times$10~cm$\times$10~cm, active detector area: 81 cm$^2$) are tiled and stacked to make a 1.6 cm thick detector module 
with an active detection area of 324 cm$^2$. The Si detectors are positioned ``above'' one hundred CZT detectors 
(each  0.5~cm$\times$2~cm$\times$2~cm) with an active detection area of 400 cm$^2$. 
The ASIC readout of the Si detectors is located at the perimeter of the Si detectors and the readout of the CZT detectors 
is located below the CZT detectors. The Si-CZT detectors are positioned above a 2~cm$\times$25~cm$\times$25~cm CsI 
active rear shield. }
\end{center}
\end{figure}

In Sect.\ \ref{HX-POL}, the concept of the hard X-ray polarimeter HX-POL is presented, a balloon-borne
experiment that uses Si and Cadmium Zinc Telluride (CZT) detectors. In Sect.\ \ref{science}, 
the science potential is discussed, and in Sect.\ \ref{discussion} the results are summarized.
\section{Description of the HX-POL experiment}
\label{HX-POL}
\subsection{Overall design, minimization of systematic errors, calibration}
HX-POL is a concept of a balloon-borne X-ray polarimeter that can measure the polarization of 
bright X-ray sources in the 50-400 keV energy range. Sketches of the gondola and the main detector assembly are shown
in Figs. \ref{gondola} and  \ref{SiCZT}, respectively. The main detector assembly is made of Si and 
CZT detectors with an active detector area of 324 cm$^2$ (Fig.\ \ref{SiCZT}). 
Photons Compton scatter in the low-Z Si and are photo-absorbed in the high-Z CZT detectors.
The polarization can be inferred by analyzing the distribution of the azimuthal scattering angles.
In addition to giving a high fraction of ``golden events'' with exactly one interaction in a Si detector and one 
in a CZT detector, the high-Z/low-Z detector combination avoids the background of neutron events which 
can mimic Compton scattered photons. 

The Si detector assembly is made of 8 layers of 2 mm thick cross-strip 
Si detectors (strip pitch: 1.4~mm), giving a total thickness of 1.6~cm. Each of the 8 layers is made of an array of four 
0.2~cm$\times$10~cm$\times$10~cm Si detectors (geometric area: 100 cm$^2$, active detector area: 81 cm $^2$).
The Si detector assembly is located above an array of 10$\times$10 CZT detectors. Each CZT detector has a 
volume of 0.5~cm$\times$2~cm$\times$2~cm and is contacted with 64 pixels (pixel pitch: 2.5~mm).
The data acquisition is triggered by a $>$20 keV hit in one of the CZT detectors with or without 
a coincident hit in the Si detectors. A coincidence unit flags Si and CZT events if the two components
triggered within a coincidence window of 2 micro-seconds. HX-POL's energy resolution includes contributions of the 
Si and CZT detectors and will be about 6\% in the 100~keV-150~keV energy region where the 
instrument sees most events.

The detector assembly resides in a 2~m high collimator assembly that shields the detectors and limits their 
field of view to a $\sim 8.4^{\circ}\times$ 8.4$^{\circ}$ region of the sky. 
Sources are observed by alternating between observations of the source (ON-observations), 
and observations of a nearby background region (OFF-observations) to enable proper background subtraction. 
The collimator has a cross section that is larger than the cross section of the detector system, resulting 
in a ``flat'' angular response and in relaxed pointing requirements. The graded collimator-shield is made 
of 4~mm lead. Thin tin and copper layers at the inside of the collimator absorb fluorescent X-ray emission 
from the lead layer. The Si and CZT detectors reside above an active shield (CsI(Na), 2~cm$\times$25~cm$\times$25~cm)
that reduces the atmospheric and internal backgrounds (e.g.\ events initiated by neutron capture in the CZT detectors) 
\cite{2006A&A...456..379G}.

Polarization measurements are notorious for systematic effects that may mimic a polarization signal.
The HX-POL design copes with systematic errors by rotating the entire collimator-detector assembly 
around the azimuthal axis at a frequency of about 0.05 Hz. After every 360$^{\circ}$ rotation the
sense of rotation is reversed. The rest of the balloon payload does not participate in the rotation. 
The collimator-detector assembly has a mass of 150 kg, and the minimization of 
torques that adversely affect the pointing of the collimator are an important design 
consideration. The azimuthal distribution of Compton scattered events can then be 
analyzed in celestial coordinates (to extract the polarization of the emission) 
and in detector coordinates (to estimate systematic errors). Analysis of the ON/OFF data can 
be used to study effects related to the orientation of the X-ray collimator relative to the 
vertical axis, and relative to the geomagnetic field.
The RHESSI experiment was rotated at 0.25 Hz, a property that was exploited for 
polarimetry measurements \cite{RHESSI,Coburn2003}.
The laboratory calibration would include tests with polarized and un-polarized beams of gamma-rays located at
different positions relative to the collimator-detector assembly. We have started first tests of a  
Si+CZT prototype experiment with polarized and un-polarized gamma-rays at Washington University.
Both tests use a 5 mCi $^{137}$Cs source. A tagged beam of polarized gamma-rays is produced by
scattering the gamma-rays off a scintillator with photomultiplier readout.  
\subsection{Si detector assembly}
\begin{figure}[t]
\begin{center}
\includegraphics[width=3.5in]{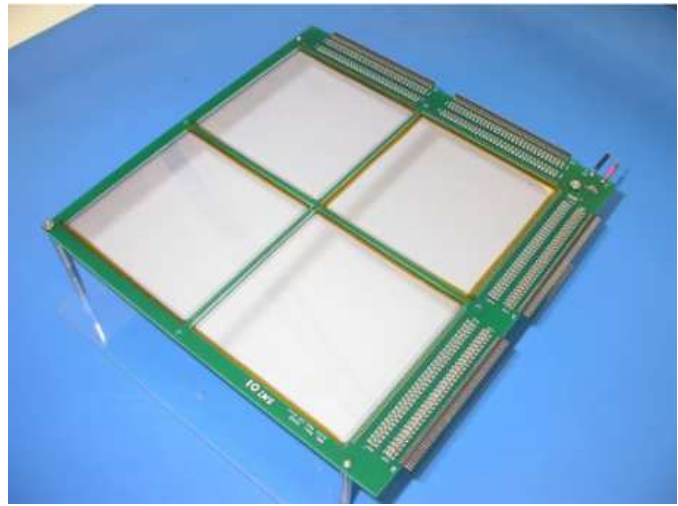}
\caption{\label{array} Photo of a 2$\times$2 array of 2 mm thick silicon strip detectors, each with a geometric area 
100 cm$^2$ and an active area of 81 cm$^2$. On the top side and on the right side, the resistors and capacitors for 
AC-coupling the crossed strips on the two sides of the detectors can be seen.}
\end{center}
\end{figure}

We chose Silicon Cross Strip Detectors (SCSDs) as the low-Z component of HX-POL. 
Si has a low atomic number of 14 which makes it a good Compton scatterer, and Si detectors have 
excellent energy and spatial resolution, even at room temperature. When cooled to temperatures 
of between -40$^{\circ}$C and -20$^{\circ}$C, they achieve energy resolutions close to those of Ge detectors. An interacting gamma ray 
deposits energy in the SCSD which leads to charge induced in the electrodes above and below the interaction.  
The amount of charge collected is directly proportional to the energy deposited in the detector.  
The strips on each side of the detector are perpendicular to each other (see the biasing components 
on two sides of the detector in Fig.\ \ref{array}).  The interaction location is defined as the 
overlapping region where strips from each electrode collected the charge \cite{Kroeger95b}.  

HX-POL uses 2~mm thick Si detectors with an area of 10$\times$10 cm$^2$ contacted with crossed 
anode and cathode strips with a strip pitch of 1.4 mm. The strips on each side of the detector 
are surrounded by a guard ring structure. The active area of each Si detector is 81 cm$^2$.
The eight layers of 2~mm thick Si detectors absorb vertically incident 50-100 keV X-rays with a 
probability of $\sim$50\%. Figure \ref{array} shows an array of four 2~mm thick Si detectors. 
The packaging of the Si-detectors in trays with eight layers of Si detectors is still under development.

\begin{figure}[t]
\begin{center}
\includegraphics[width=3.5in,height=2.8in]{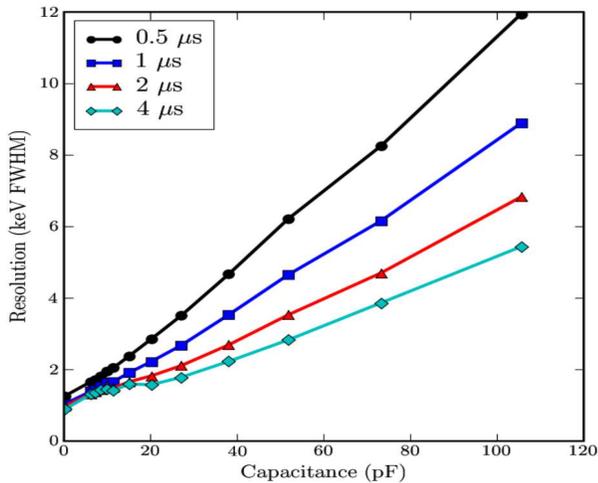}
\includegraphics[width=3.5in,height=2.8in]{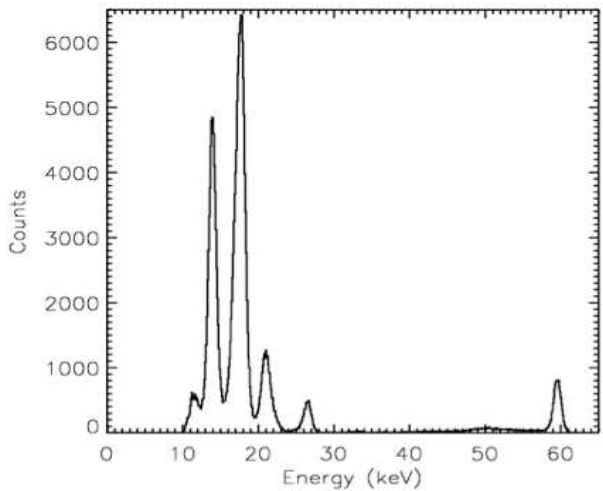}
\caption{\label{ASIC} The upper panel shows the measured room temperature resolution of the NCI-ASIC as a function 
of the input capacitance and shaping time. The resolutions refer here to Si.  The lower panel shows an $^{241}$Am 
energy spectrum obtained with one 2 mm thick, 10 cm $\times$ 10 cm Si detector, identical to the detectors
that will be used for HX-POL. The detector was cooled to -40$^{\circ}$ C for 
this measurement. The 59.5 keV FWHM resolution is 2 keV.}
\vspace*{-0.4cm}
\end{center}
\end{figure}

Both the Si and the CZT detectors are read out with the {NCI-ASIC} developed by the Brookhaven 
National Laboratory and the NRL \cite{Wulf:07,NCI-ASIC}. The ASIC combines excellent noise performance
with a sufficiently large dynamic range ($>$100) and low power dissipation (5~mW/channel).
For each of its 32 channels, the ASIC provides a low-noise preamplifier \cite{ASIC_csa,ASIC_rate}, 
a fifth order filter (shaper) with baseline stabilizer \cite{ASIC_blh}, a threshold comparator, and a peak
detector with analog memory \cite{ASIC_peak}. The ASIC properly processes charges of either polarity 
by using a design with low-noise continuous reset circuits for each polarity \cite{ASIC_rate, ASIC_reset, ASIC_gem}. 
Presently, the ASIC is used over the dynamic range from 12~keV to 2.8~MeV. Minor design changes should 
lead to a lower ASIC threshold of between 2.5~keV and 5~keV.
The noise performance measured with the ASIC is shown in the upper panel of Fig.\ \ref{ASIC}. 
For the detector capacitances relevant to HX-POL, the ASIC gives an ambient temperature dependent readout 
noise of between 1 keV and 2 keV.

The 2~mm thick Si detectors achieve excellent energy resolution. The energy resolution of one of the 
HX-POL Si detectors is shown in Fig.\ \ref{ASIC}. At 59.5 keV the energy resolution is 2 keV.
The detector was cooled to -40$^{\circ}$ C for this measurement. 
\subsection{CZT detector assembly}
CZT is the detector material of choice for the high-Z component of HX-POL. CZT has a high average 
atomic number of $\sim$50 and thus has high stopping power and a high cross section for photo-effect interactions. 
At 100 keV, a 0.5~cm thick CZT detector absorbs $\sim$99\% of incoming 100 keV X-rays, and $\sim$90\% of the first 
interactions are photo-effect interactions. Other important characteristics are operation at room temperature,
mm-spatial resolution, and good energy resolution. HX-POL will use 0.5~cm thick, 2$\times$2 cm$^2$ 
area CZT detectors contacted with 64 pixels at a pitch of 2.5 mm. Our previous results obtained with 
such detectors are described in \cite{Jung:07,Li:07} and references therein. 
In the 50~keV-100~keV energy range, photons loose only a small fraction of their energy in Compton 
interactions, and the low-energy threshold of the {HX-POL} experiment is determined by the low-energy 
threshold of the Si detectors and not by the low-energy threshold of the CZT detectors.

\begin{figure}[t]
\begin{center}
\includegraphics[width=3.5in]{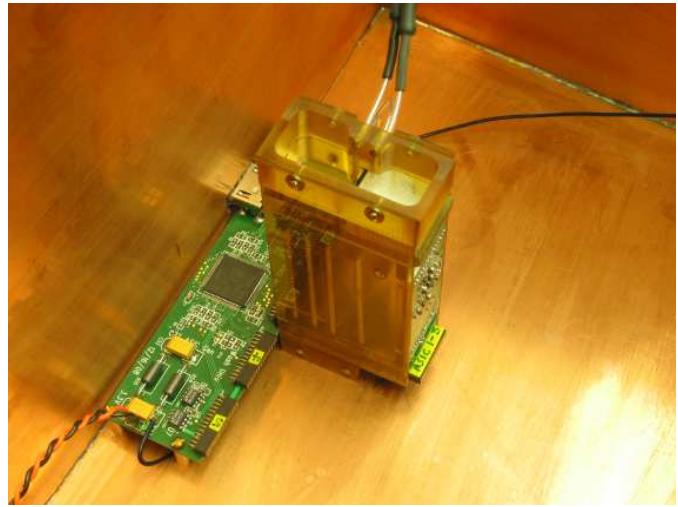}
\caption{\label{CZT} Readout electronics for the CZT array. The photograph shows two CZT detectors (each: 0.5$\times$2$\times$2 cm$^2$, 64 pixels) 
mounted in a plastic holder. The pixels are at the sides of the detectors facing the readout electronics. Inside the plastic holder reside five 
boards that hold five ASICs. Four ASICs are used to read out the pixels of the two detectors, and one 
ASIC is used to read out the cathodes. Several pairs of detectors can be daisy chained to form a ``detector chain''. 
The ``harvester board'' (visible on the left side) can read out up to five detector chains.}
\end{center}
\end{figure}
The CZT detectors are read out with the same ASICs used for the Si detectors. Fig.\ \ref{CZT} 
shows the present version of the readout system. The readout is designed to read multiple CZT 
detectors arranged in close proximity to form a ``mosaic'' of CZT detectors. 
The CZT detectors are lined up in rows and columns. Currently, the readout system can handle 5 rows 
with 5 CZT's per row, allowing for a mosaic of 25 detectors. This can easily be expanded for larger 
mosaics in the future. Each row is made up of "pair modules" daisy chained together. 
A pair-module has 2 CZT detectors (64 pixels each) mounted on a ``detector board''. 
Directly under the detector board, placed vertically, are five ``ASIC boards''. 
Four ASIC boards (each with 32 channels) read the anode pixels and a fifth ASIC board 
reads the cathode signals. The ASIC boards contain the ASICs plus 12-bit A/D converters 
and voltage regulators. They plug into a motherboard, placed horizontally, which passes digital 
signals from the ASIC boards to the "digital readout board" directly below it. 
The digital readout board contains one Field Programmable Gate Array that reads out the five ASICs 
and transmits data in a serial data stream. This configuration of boards, making up one pair module, 
fits below the footprint of the two CZT detectors, allowing for pairs of detectors to be placed side-by-side. 
Serial data from the digital readout board is daisy chained to a "harvester board". 
The harvester board reads out all five rows of pair modules and transmits the data serially to the CPU. 
Data transfers are serial LVDS (Low Voltage Differential Signaling), transmitting at a rate of 3.125 Mbits/sec. 
Dead time for an event is 125 microseconds, although we expect that this can be reduced to 70 microseconds. 

\begin{figure}[t]
\begin{center}
\includegraphics[width=3.5in]{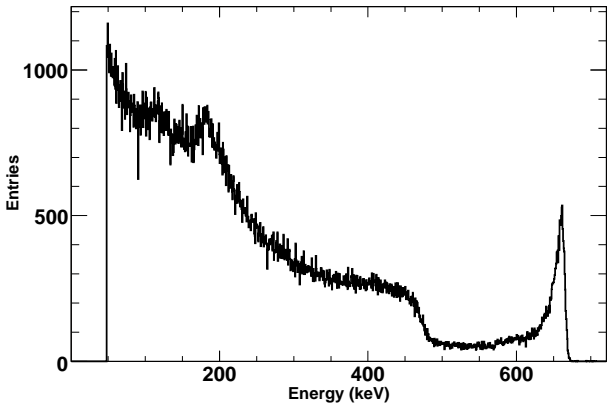}
\caption{\label{spectra} Energy spectrum taken with an 0.5 cm thick CZT detector.
The detector was operated at an energy threshold of 50 keV. The 662 keV energy 
resolution of the particular pixel shown here is 1.8\% FWHM.
}
\end{center}
\end{figure}

Fig.\ \ref{spectra} shows a $^{137}$Cs energy spectrum taken with an 0.5 cm thick CZT detector and 
a prototype version of the ASIC-readout. The detector was operated at an energy threshold of 50 keV.
The particular channel achieves an 662 keV energy resolution of 1.8\% FWHM. Note that the design of the CZT detector array 
is still evolving. Right now, the detectors are mounted in plastic holders and make contact through spring loaded 
pogo pins. An HX-POL flight module would use CZT detectors permanently mounted to ceramic chip carriers. 
We are furthermore working on reducing the size of the readout electronics. 
\section{Scientific Objectives}
\label{science}
We studied the performance of the {HX-POL} experiment with the GEANT 4 package \cite{Agos:03}
including the low-energy electromagnetic processes package GLECS \cite{Kipp:04}.
The simulations assume a Crab spectrum \cite{JoshCrab} and a ballon flight at a (zenith-angle averaged)   
atmospheric depth of 3 gr cm$^{-2}$. For this depth, the transmissivity of the residual atmosphere increases rapidly 
from 0 to 0.6 in the 20~keV to 60~keV energy range and increases slowly at higher energies.
The code simulates the energy resolution of the Si and CZT detectors and the discretization of the location 
information introduced by the strip pitch of the Si detectors and the pixel pitch of the CZT detectors. 
\begin{figure}[t]
\begin{center}
\includegraphics[width=3.5in]{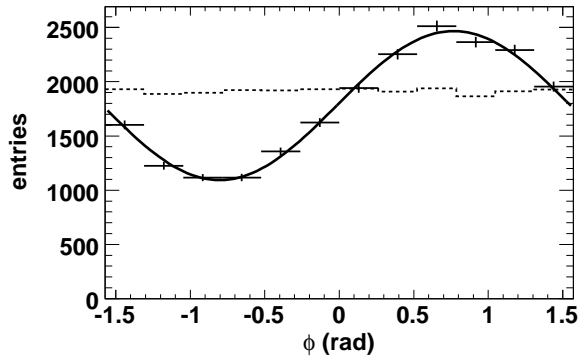}
\includegraphics[width=3.5in]{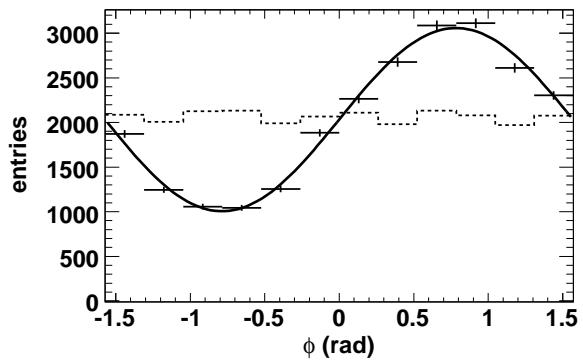}
\caption{\label{sims1} Results from the HX-POL simulations. The panels show the distributions of 
azimuthal scattering angles for events with a $>$5 keV Si detector hit and a $>$50 keV CZT detector hit (upper panel),
and for photons with a $>$5 keV hit in at least two Si detectors (lower panel) for 100\% polarized X-ray beams (solid lines) and an 
un-polarized beams (dashed lines). A source with the flux from the Crab Nebula was assumed and an ON-observation 
time of 3 hrs. The absolute numbers of counts correspond to 3 hour observations of an X-ray source with the 
flux and spectrum of that of the Crab Nebula.} 
\end{center}
\end{figure}
Two data samples were simulated: one with a polarization signature (the ``ON-data sample'') and one without (the ''OFF data sample'').
The analysis of the simulated data uses only information that will also be available experimentally: the signal amplitudes 
measured in the Si strips and in the CZT pixels. If more than two interactions are recorded (taking into account the spatial 
resolution of the detectors), only the two highest energy deposits are used to determine the azimuthal scattering angle.

\begin{figure}[t]
\begin{center}
\includegraphics[width=3.5in]{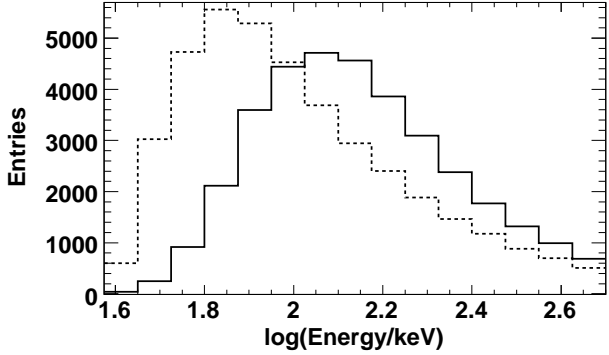}
\caption{\label{sims2} The panel shows the energy distribution of photons producing Si-CZT events (solid line) and
Si-Si events (dashed line).}
\end{center}
\end{figure}

Figures \ref{sims1} and \ref{sims2} show results for 100\% polarized and un-polarized X-ray beams (no backgrounds). 
The absolute number of events correspond to a 3 hr ON-observation of a source with the flux and spectrum 
of that of the Crab Nebula \cite{JoshCrab} (the emission from the Crab Nebula is of course not polarized to 100\%).
The upper panel of Fig.\ \ref{sims1} shows the distribution of azimuthal scattering angles for ``Si-CZT'' events 
where one Si detector records a hit (most likely from a Compton interaction) and one CZT detector pixel records 
a hit (most likely from a photo-effect interaction). The lower panel of Fig.\ \ref{sims1} shows the same for ``Si-Si'' 
events with hits in two or more Si detectors. The analysis of Si-Si events discards events in which the two highest 
energy depositions are less than than 3.2 pixel pitches away from each other when projected onto the plane of the detectors.
This cut leads to a rather even distribution of detected azimuthal scattering angles.  The events that do not pass this 
cut also hold information and could be used in a more sophisticated analysis.

The modulation factor is defined as
\begin{equation}
\mu\,=\,\frac{C_{\rm max}-C_{\rm min}}{C_{\rm max}+C_{\rm min}}
\end{equation}
where $C_{\rm max}$ and $C_{\rm min}$ refer to the maximum and minimum counts detected for different azimuthal scattering angles.
For the Si-CZT and the Si-Si events the modulation factors are $\mu_{\rm Si-CZT}\,=$ 0.4, and $\mu_{\rm Si-Si}\,=$ 0.5, respectively.
The simulations can be used to determine the minimum detectable polarizations (MDP). 
We compute the 99\% confidence level MDP with a modified version of the Equation (10)
in \cite{Weisskopf} that accounts for the statistical errors of the OFF-data:
\begin{equation}
{\rm MDP}\,=\,\frac{4.29}{\mu R_{\rm src}}\sqrt{\frac{R_{\rm src}+2\,R_{\rm bg}}{T}} 
\end{equation}
where $R_{\rm src}$ is the total source counting rate,  $R_{\rm bg}$ is the total background 
counting rate and $T$ is the ON integration time (assumed to equal the OFF integration time).
\begin{table}[!t]
\renewcommand{\arraystretch}{1.3}
\caption{HX-POL Minimum Detectable Polarization (MDP).}
\label{tab1}
\centering
\begin{tabular}{|c||c|c|c|c|}
\hline
Events & $\mu$ & $R_{\rm src}\,\left[\rm Hz\right]$ & $R_{\rm bg}\,\left[\rm Hz\right]$ & MDP \,$\left[\rm \%\right]$\\ \hline \hline
Si-CZT & 0.40 & 2.0 & 0.13 & 7.7\\
Si-Si  & 0.50 & 2.3 & 0.14 & 5.8 \\
All    & 0.46 & 4.3 & 0.27 & 4.7 \\ \hline
\end{tabular}
\end{table}

Table \ref{tab1} lists the values $R_{\rm src}$, $R_{\rm bg}$, and MDP for the Si-CZT events, the Si-Si events, and 
for both types of events. We assume interspersed ON-OFF observations with a ON and OFF integration time
of 3 hrs each. The background was adopted from the balloon results reported by Sch\"onfelder et 
al.\ \cite{Scho:77}. The prediction of the sensitivity of the Si-CZT events should be robust. 
The sensitivity of the Si-Si events will be poorer than stated in Table \ref{tab1}, as we did 
not attempt to model neutrons. Elastically scattering neutrons can mimic Compton-events. 
Figure \ref{sims2} shows the distribution of the energy of the Si-CZT events and the Si-Si events. 
Most Si-CZT events have energies between 60 keV to 400 keV. Most Si-Si events have energies between 
50 keV and 300 keV. 

\begin{figure}[t]
\begin{center}
\includegraphics[width=3.5in]{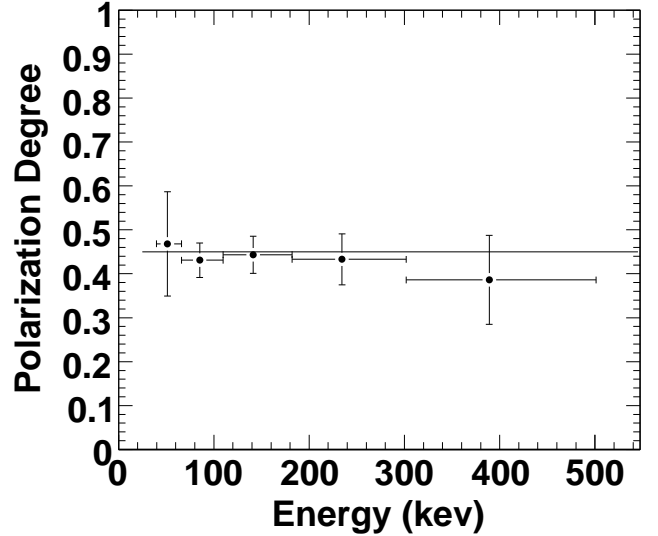}
\caption{\label{sims3} The graph shows the results of a simulated observation of the Crab Nebula (3 hours ON, 3 hours OFF).
The polarization degree can be measured in several independent energy bins.}
\end{center}
\end{figure}

Figure \ref{sims3} shows the measured polarization degree for simulated observations of the Crab Nebula. For this graph, 
only Si-CZT events were used. For a balloon launch from Fort Sumner (NM) or Palestine (TX), the Crab Nebula can be 
observed for 6 hours at zenith angles smaller than 40$^{\circ}$. We thus assumed an observation with 3 hours of ON-time 
and 3 hours of OFF-time. As before, we use a zenith angle averaged residual atmosphere of 3 gr cm$^{-2}$. 
The graph shows that HX-POL will be able to measure the polarization of the 
Crab Nebula in several independent energy bins. 

Additional observation targets for a one-day balloon-flight are black hole binaries and accreting neutron stars. 
The binary black holes Cygnus X-1 and GRS 1915+105 are prime targets because of their X-ray brightness. 
Cygnus X-1 is an exceptional source as it shows persistent flaring activity. 
In the HEAO-4 survey it was {\it the} brightest $>80$ keV source \cite{Levi:84}. 
The source GRS 1915+105 is well known for frequent flaring epochs with a stunning diversity of flux and 
spectral patterns \cite{McCl:06}. Binary black holes exhibit different emission 
states (thermal, hard, and steep power law) that depend on the structure of the accretion disk and the accretion disk 
corona, and on the relative importance of a relativistic outflow (jet) \cite{Remi:06}. 
Synchrotron emission and Compton scatterings could both lead to polarization degrees that are observable with HX-POL. 
Bright accreting neutron stars are  Hercules X-1 and Cen X-3. Owing to the emission mechanism and to the 
propagation of the X-rays through the highly magnetized plasma surrounding the neutron stars, the X-rays 
from these objects might exhibit polarization degrees of several 10\% (e.g. \cite{Kii:87,Mesz:88}). 

On a longer 15-30 day long balloon flight, HX-POL would be able to scrutinize the polarization in strong galactic sources
to a level of a few percent. Deep observations could be used to measure the energy dependence of the polarization of the
X-ray emission from binary black holes. These measurements would constrain the dominant emission process as a function of energy.
In high magnetic-field pulsars and magnetars, photon splitting (a higher order QED effect) could produce strong 
energy-dependent polarization leading up to a high-energy cut-off \cite{Bari:01,Hard:97}. This effect cannot be
observed in terrestrial laboratories and would provide a powerful test of QED in extreme conditions.

A longer balloon flight would have a realistic chance to measure the polarization of the hard X-rays 
from extragalactic sources, e.g. from ``extreme synchrotron blazars'', mass-accreting supermassive black holes 
with highly relativistic jets pointing at the observer \cite{Cost:01}. Their synchrotron emission extends into the very hard 
X-ray regime. For example, for Mrk 501, the BeppoSAX mission measured a spectral energy distribution with a 
``low-energy'' component peaking in the $>$100 keV energy range \cite{Pian:98}. The high-energy component, presumably 
inverse Compton emission, peaks in the GeV/TeV energy range \cite{Kraw:04}. Extreme synchrotron blazars show long lasting 
periods (3 months or more) of intense flaring activity. A positive detection of a high polarization degree 
would show that the magnetic fields in AGN jets are well ordered in the ``blazar zone'' $\sim$1 pc 
away from the black hole, and would add an important constraint to models of jet formation.

\section{Summary and conclusions}
\label{discussion}
This paper describes the concept of a balloon-borne hard X-ray polarimeter. 
The detector module combines a low-Z Compton scatterer (Si) with a high-Z photoeffect absorber (CZT).
The detector combination achieves excellent polarization sensitivity in the 50 keV to 400 keV energy band.
Based on 3+3 hrs of ON-OFF observations during a one-day balloon flight, 
the polarization of strong Crab-like sources can be measured for polarization 
degrees well below 10\%. On a 15-30 day balloon flight, polarization degrees of a few percent can be measured. 
The 50-400 keV energy band is well suited for a balloon-borne experiment: atmospheric absorption hinders 
observations below 20 keV and observations above 400 keV have to deal with very low photon fluxes and require 
relatively large experiments and/or long integration times, both better suited for space borne experiments. 
One of the strengths of the Si/CZT detector combination is the excellent energy resolution that 
makes it possible to measure the energy dependence of the polarization properties.
Compared to experiments that use Si detectors only, the Si/CZT combination avoids the neutron
backgrounds that plague  experiments exclusively made of low-Z detectors. Compared to experiments 
that use CZT only, the Si/CZT combination achieves a lower energy threshold. 
In CZT detectors, photoeffect interactions dominate over Compton interactions up to 
energies of $\sim$250 keV; furthermore, $<$100 keV photons are absorbed quickly in 
CZT, making it necessary to use CZT detectors with very small pixel pitches.
\section*{Acknowledgment}
The Washington University group acknowledges support by NASA (grant NNX07AH37G). 

\end{document}